# Giant tunability of the two-dimensional electron gas at the interface of γ-Al$_2$O$_3$/SrTiO$_3$

*Wei Niu,[†,‡] Yu Zhang,[†] Yulin Gan,[†] Dennis V. Christensen,[†] Merlin V. Soosten,[†] Eduardo J. Garcia-Suarez,[§] Anders Riisager,[§] Xuefeng Wang,*[‡] Yongbing Xu,[‡] Rong Zhang,[‡] Nini Pryds[†] and Yunzhong Chen*[†]*

[†]Department of Energy Conversion and Storage, Technical University of Denmark, Risø Campus, 4000 Roskilde, Denmark

[‡]National Laboratory of Solid State Microstructures, School of Electronic Science and Engineering, Nanjing University, 210093 Nanjing, China

[§]Center for Catalysis and Sustainable Chemistry, Department of Chemistry, Technical University of Denmark, 2800 Lyngby, Denmark

ABSTRACT:

Two-dimensional electron gases (2DEGs) formed at the interface between two oxide insulators provide a rich platform for the next generation of electronic devices. However, their high carrier density makes it rather challenging to control the interface properties under a low electric field through a dielectric solid insulator, i.e. in the configuration of conventional field-effect transistors. To surpass this long-standing limit, we used ionic liquids as the dielectric layer for electrostatic gating of oxide interfaces in an electric double layer transistor (EDLT) configuration. Herein, we reported giant tunability of the physical properties of 2DEGs at the spinel/perovskite interface of γ-$Al_2O_3$/$SrTiO_3$ (GAO/STO). By modulating the carrier density thus the band filling with ionic-liquid gating, the system experiences a Lifshitz transition at a critical carrier density of $3.0\times10^{13}$ cm$^{-2}$, where a remarkably strong enhancement of Rashba spin-orbit interaction and an emergence of Kondo effect at low temperatures are observed. Moreover, as the carrier concentration depletes with decreasing gating voltage, the electron mobility is enhanced by more than 6 times in magnitude, leading to the observation of clear quantum oscillations. The great tunability of GAO/STO interface by EDLT gating not only shows promise for design of oxide devices with on-demand properties, but also sheds new light on the electronic structure of 2DEG at the non-isostructural spinel/perovskite interface.

The metallic interface between two insulating oxides, a two-dimensional electron gas (2DEG) in nature, provides a rich platform for exploring new fundamental phenomena and device applications.[1-3] In particular, as silicon is the foundation of semiconductor technology, the perovskite oxide insulator $SrTiO_3$ (STO) is the base material for oxide electronics. Different from conventional semiconductor 2DEGs, the STO-based 2DEG originates from Ti $3d$ orbits ($d_{xy}$, $d_{xz}$ and $d_{yz}$)[4] and exhibit intriguing properties, such as two-dimensional superconductivity,[5] signature of magnetism,[6] high carrier mobility,[7] nanoscale-controlled insulator-metal transition,[8] 2D quantum oscillations,[9] large thermopower modulation[10] and sensitivity to light illumination[11]. These properties depend strongly on the carrier density due to electron-electron correlations. Therefore, extensive efforts have been made to tune the interface properties by electrostatic gating. Generally, the modulation of these novel properties of oxide 2DEGs is performed in conventional field effect configuration where the STO substrate is used as the dielectric insulator. However, this requires high voltages of tens to hundreds of volts to achieve a sizable field effect.[12-14] For example, the Lifshitz transition where the $d_{xz}$ and $d_{yz}$ bands start to become occupied was observed to emerge with the back gate ranging from -50 V to 450 V, near a critical carrier density of $1.7 \times 10^{13}$ cm$^{-2}$ in $LaAlO_3/SrTiO_3$ (LAO/STO).[15] Furthermore, applying the back gate between -200 V and 200 V, Rashba spin-orbit interaction (SOI) was effectively tuned in the isostructural 2DEG systems, such as LAO/STO[16] and $LaVO_3/SrTiO_3$.[17] Different from the studies with conventional electric field effects, electric double layer transistor (EDLT) with an ionic liquid (IL) as the dielectric layer provides a more powerful means to tune the carrier density as high as ~$10^{15}$ cm$^{-2}$ with only a few volts.[18, 19] Such ionic gating effect has resulted in the observation of field-induced superconductivity in insulating STO [20], as well as Kondo effect in LAO/STO.[21, 22]

Besides the intensively investigated isostructural perovskite/perovskite interfaces, recently, a new 2DEG was discovered at the non-isostructural spinel/perovskite oxide interface of γ-Al$_2$O$_3$/SrTiO$_3$ (GAO/STO).[2, 14, 23-27] Compared to the mostly investigated LAO/STO interface, the GAO/STO system has better lattice match, much higher electron mobility and does not contain rare earth elements, and therefore is more attractive for application in high-mobility oxide devices. However, the high-mobility GAO/STO interface is generally accompanied by a high carrier density (on the order of 10$^{14}$ cm$^{-2}$), which is approximately one order larger than that of LAO/STO (on the order of 10$^{13}$ cm$^{-2}$).[26] This makes it more challenging to modulate the physical properties of GAO/STO by the conventional configuration.[14] Therefore, it becomes extremely interesting to investigate the GAO/STO electrostatically gated by ionic liquids. Nevertheless, so far the IL-assisted EDLT at oxide interfaces remains at its early stage, which is largely limited to the isostructural LAO/STO heterostructures. Whether the remarkable tunability of interface properties in LAO/STO can be also achieved for the non-isostructural GAO/STO remain opens. Moreover, the electronic structure underlying the novel properties of LAO/STO system is well understood, including the evolution of the Fermi level under electrostatic gating effect. Nevertheless, little is known to the GAO/STO system. Notably, it was recently reported by resonant soft X-ray linear dichroism,[23] that GAO/STO could exhibit an unusual orbital symmetry different from that of the LAO/STO. This makes it further interesting to study the gating effect of the spinel-perovskite heterostructure. Herein, for the first time, we applied the EDLT technique to the non-isostructural GAO/STO oxide interface. By continuously tune the carrier concentration of the interface 2DEG, the system undergoes a series of remarkable transitions from Lifshitz transition, to enhanced SOI, as well as the appearance of Kondo effect. Besides the large tunability of the interface states, our results also shed new light on the electronic structure on the non-isostructural GAO/STO interface.

The GAO/STO EDLT device, as illustrated in Figure 1a, was fabricated on a Hall-bar 2DEG[28, 29]. The Hall bar device was fabricated by initially depositing amorphous LaMnO$_3$ (a-LMO) layer (50 nm) as a hard mask layer. Optical lithography and selective wet etching processes were subsequently performed to create patterned STO substrates for the deposition of GAO (see also the Supporting Information). The width of Hall bar is 50 μm and the length between two voltage probes is 500 μm. A drop of ionic liquid, 1-ethyl-3-methylimidazolium-bis(trifluoromethanesulfonyl) amide (EMI-TFSI) with poly(styrene-block-methylmethacrylate-block-styrene) (PS-PMMA-PS) which was made into the form of gel with a freezing point of 230 K, was chosen as the gate dielectric material due to the high capacitance.[30] The gate voltage is applied through a Pt electrode to the ionic liquid. Note that although the Pt contact to LAO/STO results in resistance hysteresis,[31] we didn't observed any hysteresis during our gating measurement. In this paper, the data were collected from a typical device with a GAO film thickness, $t$, of 2.5 unit cells (uc) where the interface is metallic with a relative high electron mobility. Similar measurements were performed on several devices with $2\ uc \leq t \leq 2.5\ uc$. All data show nice consistency and reproducibility ruling out any possible electrochemical reaction[32] or aging effect[33]. Notably, the capping layer of GAO film could play a similar role as a boron nitride buffer to prevent the ionic-liquid-induced electrochemical reaction on STO surface.[32] Figure 1b shows an atomic force microscopy (AFM) image of the 2.5 uc GAO film grown on the patterned STO substrate. An atomically smooth surface with clear terraces is observed, indicating the high quality of both the patterned substrate and the grown film.[29] Figure 1c shows the temperature ($T$)-dependent sheet resistance ($R_s$) and carrier density ($n_s$) of this device before dropping the ionic liquid. The patterned GAO/STO is metallic as reported previously.[29] The carrier density, $n_s$, is approximately $2.2 \times 10^{13}$ cm$^{-2}$ at room temperature, and is nearly independent on temperature in the whole temperature range of 2-300 K. The electron mobility, $\mu$, at 2 K is around 1,800 cm$^2$V$^{-1}$s$^{-1}$. Notably, compared to the $n_s = 3.7 \times 10^{14}$ cm$^{-2}$ of unpatterned GAO/STO system with the same thickness, the carrier density of the patterned 2DEG is suppressed significantly. This is probably due to the presence of LaMnO$_3$ (which is crystallized from the a-LMO hard mask during the high temperature deposition) in the structure which could change significantly the oxygen exchange dynamics across the interface at high temperatures.[26, 29]

The work mechanism of our EDLT device is illustrated in the insets of Figure 2a: By applying a positive gate voltage to the Pt electrode, the cations and anions in the ionic liquid move towards the sample and the gate electrode, respectively, forming an electric double layer at the liquid-solid interface (under the negative gate voltage, the cations and anions in the ionic liquid move towards the Pt electrode and the GAO/STO interface, respectively). This electric double layer works as a nanogap supercapacitor tuning the carrier density as well as the Fermi level in the electronic structure. When applying a positive gate voltage, more electrons are accumulated at the interface. However, applying negative voltage to the gate electrode, electrons are depleted. Figure 2a displays the $R_s$ as a function of temperature at various $V_g$. All the $R_s$-$T$ curves show overall metallic behaviors, namely, the sheet resistances decrease upon cooling. In the high temperature regime such as $T$=200 K, the sheet resistance decreases monotonously from ~20,000 $\Omega/\square$ to ~7,000 $\Omega/\square$ as the $V_g$ increases from -1.5 V to 3 V, indicating the accumulation of carriers upon increasing the electrostatic gate potential as expected. However, at low temperature region, a remarkable upturn in the $R_s$ occurs. Particularly, at $V_g \geq 2$ V, $R_s$ reaches firstly a minimum at a certain temperature and then increases until $T$ decreases to 2 K. Similar electrostatic modulated resistance minimum have also been observed in LAO/STO system.[21, 34] To make the resistance minimum more obvious, in Figure 2b, we show the offset sheet resistance [$R_s(T)$-$R_s$(2 K)] at low temperature between 2 K to 40 K. The abnormal resistance minimum could result from either the weak localization[35, 36], or the Kondo effect[37] as reported previously. However, the possible weak localization mechanism can be ruled out here based on our magnetoresistance measurements as discussed in the latter part.[35] We, therefore, account the resistance minimum comes from the Kondo effect, which normally arises from the exchange interaction between itinerant conduction electrons and localized spin centers.[36] The observed resistance minimum can be described well by a simple Kondo model[21]

$$R(T) = R_0 + aT^b + R_{K,0} \left( \frac{1}{1+(2^{1/s}-1)(T/T_K)^2} \right)^s \tag{1}$$

where $R_0$ is the residual resistance, the second power-law term are the contributions from electron-electron and electron-phonon interactions. The last term explains the Kondo effect contribution. $T_K$ is the Kondo temperature that characterizes the strength of the Kondo effect and the parameter $s$ is fixed at 0.225 for STO-based 2DEG system.[21, 22] Good fitting results were obtained (Supporting Information Figure S3) with Kondo temperatures, $T_K$, of 51.1 K and 52.8 K at $V_g$=2 V and 3V, respectively. The positive correlation between the resistance upturn and the enhanced carrier density is consistent with the characteristic of the Kondo effect.[21] Localized electrons, particularly with a $d_{xy}$ orbital occupation, could act as magnetic centers for STO-based heterostructures,[4, 22] including GAO/STO.[38] But the emergence of the Kondo effect at $V_g$ = 2 V, i.e. at a high carrier density of $n_s \sim 3\times10^{13}$ cm$^{-2}$ is unconventional since the magnetic scattering due to the presence of the magnetic impurities is expected to be stronger at a lower carrier density. The observed appearance of the Kondo effect in our interface as a function of an applied electric field, is quite similar to the case of the ionic gated STO surface,[22] and points to the emergence of magnetic interactions between electrons in STO due to electron-electron correlations rather than the presence of dopants. However, the precise threshold density for the emergence of the Kondo effect of these two system is different, i.e. the threshold density of the ionic gated bare STO is 9.2×10$^{13}$ cm$^{-2}$, much higher than that of GAO/STO interface (3.0×10$^{13}$ cm$^{-2}$).

Figures 3a and b show the Hall resistance, $R_{xy}$, as a function of magnetic field (up to 16 T) for -1.5 V ≤ $V_g$ ≤ 2 V and  2 V < $V_g$ ≤ 3 V, respectively. Starting from the lowest gate voltage $V_g$= -1.5 V and up to a transition value of $V_g$= 2 V, $R_{xy}$ is linear and the Hall coefficient, $R_H$, decreases with increasing $V_g$, as shown in Figure 3a, which indicates a single band conductivity. The carrier density can be deduced using $n_s$= -1/$R_H$e, which gives $n_s$=

$1.3×10^{13}$ cm$^{-2}$ at $V_g$=-1.5 V and $n_s$= $3.0×10^{13}$ cm$^{-2}$ at $V_g$=2 V. However, as $V_g$ > 2 V (Figure 3b) nonlinear Hall effect appears, which is likely due to the presence of the multiband transport carriers as observed in perovskite/perovskite 2DEG systems.[39] We, therefore, use the two-band model to fit the results (details in Supporting Information Figure S4), where n$_1$ and n$_2$ represent the electron densities for the two different bands. Figure 3c summarizes the extracted carrier density as a function of gate voltage. As we can see, $n_1$ contributes exclusively to the conduction at $V_g$< 2 V, and it increases linearly with increasing $V_g$. This is consistent with the single-band-type carrier conduction model. At $V_g$>2 V, $n_2$ emerges and increases with the increase of the $V_g$. However, $n_1$ begins to decrease unexpected. On the other hand, the total carrier densities, $n_{total}$=$n_1$+$n_2$, increases (from 1.3 to $3.5×10^{13}$ cm$^{-2}$) within gate voltage range from -1.5 V to 3 V.

For the STO-based 2DEGs, it is established that the thin sheet conducting electrons are confined strongly within a few-nanometers at the interface on the STO side.[40] The interfacial electrons originate from the electronic shell of the Ti 3$d$ orbits ($d_{xy}$, $d_{xz}$ and $d_{yz}$).[4] Subsequently, the spin and momentum of electrons in 2DEGs are entangled and complicated sub-bands are formed due to the strongly anisotropic nature of $d$ orbits and the quantum confinement.[41] Generally, the band bending at the interface lifts the degeneracy of Ti $t_{2g}$ levels and produces an energy splitting between the different levels: $d_{xy}$ electrons occupy the lowest energy states of the 2DEG; they account for most of the charge in the system. By contrast, $d_{xz}$/$d_{yz}$ electrons amount to only a small fraction of the population but occupy further higher energy states. With this universal electronic structure in mind, [4, 42] we therefore, propose that: at $V_g$< 2 V, electrons ($n_1$) locate at the lowest $d_{xy}$ levels (as depicted in Figure 3d). Upon the population of electron as increasing $V_g$, the Fermi level is lifted across the bottom of $d_{xz}$/$d_{yz}$ bands, the system undergoes a remarkable Lifshitz transition at a critical carrier density, [15, 39] $n_L$~ $3.0×10^{13}$ cm$^{-2}$ (Figure 3e), above which the $d_{xz}$/$d_{yz}$ bands start to populate (corresponding to the appearance of experimental $n_2$) (Figure 3f).

Notably, the critical carrier density, $n_L \sim 3.0 \times 10^{13}$ cm$^{-2}$, for the Lifshitz transition of GAO/STO is higher than most of the reports for LAO/STO where the transition is achieved with a back gate. However, it is remarkably close to the $n_L$ reported previously for top-gated LAO/STO systems.[39] Nontrivially, the unexpected decrease in $n_1$ upon increasing $V_g$ i.e. the depletion in the occupation of $d_{xy}$ states at 2 V < $V_g \leq$ 3 V was also observed in the top-gated LAO/STO systems.[39] Such a decrease of $d_{xy}$ electrons is inconsistent with a model requiring a fixed electronic band structure, as raising the Fermi energy should always increase the number of available conduction states.[39] Therefore, our data also indicate that the underlying band structure of GAO/STO is not fixed under ionic gating. Instead, carriers in $d_{xy}$ and $d_{xz}/d_{yz}$ bands will redistribute and the band structure evolves with the electrostatic gating effect. Based on Schrödinger-Poisson calculations proposed in Ref. 39, the $d_{xz}/d_{yz}$ energy increases faster than the energy of the $d_{xy}$ states and the potential well at the Lifshitz point is narrow with the top-gate configuration. In addition, a positive top-gate voltage enhances the confining potential gradient, resulting in a larger splitting of the bands as depicted in Figure 3f. Therefore, once the $d_{xz}/d_{yz}$ electrons are populated, band occupations are influenced and $d_{xy}$ electrons are suppressed, strongly highlight the electron-electron interactions at complex oxide interfaces as reported recently.[39] This is consistent with the conclusion proposed by Maniv *et al*. that electronic interactions cause a competition between the occupancies of different bands.[43] Notably, the Kondo effect emerges as $d_{xy}$ electrons decreases upon increasing $V_g$, although it appears unusually as the total carrier density increases. Shortly, the IL gating effectively accumulate/dissipate carriers at the interface by modifying deeply the band structure of the 2DEG as well as electron-electron interactions.

Besides carrier densities, Rashba spin-orbit interaction (SOI) arising from the interfacial breaking of inversion symmetry can also be modulated by the external electric field, particularly at the crossing point of $d_{xy}$ and $d_{xz}/d_{yz}$ orbitals. This could be useful to control the spin precession in spintronics devices.[16] The significant tuning of Ti 3$d$ bands as revealed in Fig.3 may lead to unique spin-orbit textures, particular at $V_g$=2V, i.e. in proximity to the Lifshitz transition. This inspires us to further study the gate-dependent SOI at the GAO/STO interface. Figure 4a shows the magnetoresistance (MR) as a function of magnetic field measured at 2 K with various applied $V_g$. Notably, no negative magnetoresistance or weak localization effect was observed in our measurements. This rules out the possibility of a weak localization effect that leads to the resistance minimum discussed in Figure 2b. Positive MR curves were observed in the whole field region of (-1.5 V$\leq V_g \leq$ 3 V). Remarkably, the MR exhibits sharp cusps at low magnetic fields for Vg $\geq$0 V, which is the characteristic feature of the weak antilocalization (WAL) effect originating from the interference of the quantum coherent electronic waves in the presence of SOI.[44-46] In Figure 4b, we plot the $\Delta\sigma = \sigma(\mu_0 H) - \sigma(0)$ at different $V_g$, where the WAL effect at low magnetic fields becomes more visible as Vg $\geq$0 V.

To understand the SOI tuned by the IL-assisted gating, we analyzed the observed WAL effect using the Maekawa-Fukuyama (MF) localization theory:[16, 47]

$$\Delta\sigma(H) = \frac{e^2}{\pi h}[\Psi(\frac{H}{H_i + H_{so}}) + \frac{1}{2\sqrt{1-\gamma^2}}\Psi(\frac{H}{H_i + H_{so}(1+\sqrt{1-\gamma^2})}) - \frac{1}{2\sqrt{1-\gamma^2}}\Psi(\frac{H}{H_i + H_{so}(1-\sqrt{1-\gamma^2})})] \quad (2)$$

Here, the function $\Psi$ is defined as $\Psi(x) = \ln(x) + \psi(1/2 + 1/x)$, where $\psi(x)$ is the digamma function. $H_i$, $H_{so}$ and $\gamma$ are parameters indicate the characteristic inelastic scattering field, spin-orbit interaction field and Zeeman correction term, respectively. The magnitude of the characteristic spin-orbit interaction field ($\mu_0 H_{so}$) reflects the SOI strength. Figure 4c displays the fitting results of the experimental data according to **Equation 2**. The parameters $\mu_0 H_i$ and $\mu_0 H_{so}$ extracted from fitting are shown in Figure 4d, which shed light on the modulation of SOI by the electric field. The values of $\mu_0 H_{so}$ are in the same range as the ones reported previously for other STO-based 2DEGs.[16, 48, 49] For negative gate voltages, the inelastic scattering field is larger than the spin-orbit interaction field, indicating that the orbit effect of magnetic field dominates compared with the effect of spin-orbit interaction. While at $V_g \geq 0$ V, the spin-orbit interaction field is larger than the inelastic scattering field and increases significantly at $V_g > 2V$. This further confirms the scenario of band structure discussed above in Figure 3d-f, since significant SOI enhancement has been theoretically predicted and experimental proved to occur at the $d_{xy}$-$d_{xz}$/$d_{yz}$ crossing region due to the orbital hybridization.[17, 42, 50, 51] At negative $V_g$ regions of single-band at lower density, electrons only occupy the $d_{xy}$ sub-bands without the crossing between $d_{xy}$ and $d_{xz}$/$d_{yz}$, the SOI does not dominate. However, at $V_g > 2V$, i.e. after the Lifshitz transition, the $d_{xy}$ and $d_{xz}$/$d_{yz}$ sub-bands begin to cross at Fermi level and subsequently have the biggest crossing area around Fermi level ($V_g = 2.5V$). The demonstration of the field control of Rashba SOI in GAO/STO EDLTs presents not only a step towards realizing spintronics devices based on IL gating, but also shed some light on the underlying mechanism of the Rashba SOI and $3d$ band filling of GAO/STO interface modulated by IL gating.

Finally, as shown in Figure 4a, we found clear Shubnikov-de Hass (SdH) oscillations in the MR curves for $V_g = -1$ V and $-1.5V$, which is a characteristic behavior for high-mobility high quality materials.[9, 26] To further demonstrate the modulation of mobility by IL-assisted electrostatic effect, Figure 5a shows the mobility as a function of $V_g$. The mobility is enhanced by depleting the carriers with decreasing the gate voltages.[34] Notably, at $V_g < 0$ V, the electron mobility is increased larger than 2000 cm$^2$V$^{-1}$s$^{-1}$ at 2 K. Figure 5b shows SdH oscillations recorded at 2 K as a function of the inverse magnetic field $1/\mu_0 H$ after removing a smooth background. The inset of Figure 5b shows the position of oscillation peaks in $1/\mu_0 H$ versus the effective Landau level, of which the fitted lines indicate the frequency of SdH oscillations. The oscillation frequency decreases as carrier density decreases, which is consistent with a previous result for doped STO.[52] The carrier density determined from Hall measurements

($n_{\text{Hall\_-1.5V}}$ = 1.3×10$^{13}$ cm$^{-2}$, and $n_{\text{Hall\_-1V}}$ = 1.7×10$^{13}$ cm$^{-2}$ for $V_g$=-1.5 V and -1V, respectively) and SdH oscillations ($n_{\text{SdH\_-1.5V}}$ = 3.2×10$^{12}$ cm$^{-2}$, $n_{\text{SdH\_-1V}}$ = 4.8×10$^{12}$ cm$^{-2}$) differ by a factor of 3-4 in the present experiments. Similar behaviors have been observed in previous reports for unpatterned samples.[26, 29, 43, 53] This discrepancy is either due to the fact that a fraction of carriers do not satisfy condition for the SdH oscillation but nevertheless contribute to the Hall signal, or due to the presence of multiple conduction channels with same carrier mobility. Since our SdH oscillations are observed in the single band conduction region ($V_g$= -1.5 V and -1V) with only $d_{xy}$ electrons populated to the conduction, we assume that the GAO/STO interface consists of a single quantum well with approximately 4–fold degeneracy, such as two $d_{xy}$ conduction channels with spin splitting.[53, 54] Overall, by IL-assisted field effect, the mobility can be highly enhanced leading to quantum oscillations.

In summary, we have demonstrated the tuning of the 2DEG at the GAO/STO interface by electric-double-layer gating. A dramatic tuning of the electronic structure of GAO/STO is realized in which the electron gas undergoes a number of remarkable transitions from Lifshitz transition to the occurrence of Kondo effect, as well as a large enhancement in the Rashba SOI. Furthermore, EDLT at the GAO/STO interface also shows an important avenue to construct high-mobility oxide interface where nontrivial quantum phenomena could be explored. The present findings suggest the IL-assisted field effect of non-isostructural oxide interface of GAO/STO is promising for novel all-oxide devices.

ASSOCIATED CONTENT

**Supporting Information**

Sample growth, device and ionic liquid gel fabrication, figures of EDLT device, leak current, Kondo effect and two-band model (PDF). This material is available free of charge via the Internet at http://pubs.acs.org.

## AUTHOR INFORMATION

**Corresponding Author**

*E-mail: (Y. C.) yunc@dtu.dk
*E-mail: (X. W.) xfwang@nju.edu.cn

**Author Contributions**

The manuscript was written through contributions of all authors. All authors have given approval to the final version of the manuscript.

**Notes**

The authors declare no competing financial interest.

## ACKNOWLEDGMENT

We thank the technical help from Jørgen Geyti and the scientific discussion with Søren Linderoth. Wei Niu thanks the support by China Scholarship Council. X.F.W. acknowledges the financial support from the National Key Projects for Basic Research of China under Grant No. 2014CB921103, the National Natural Science Foundation of China under Grant No. U1732159, U1732273 and the Collaborative Innovation Center of Solid-State Lighting and Energy-Saving Electronics. Y.B.X. acknowledges the financial support from the Natural Science Foundation of Jiangsu Province of China under Grant No. BK20140054.

# Figures

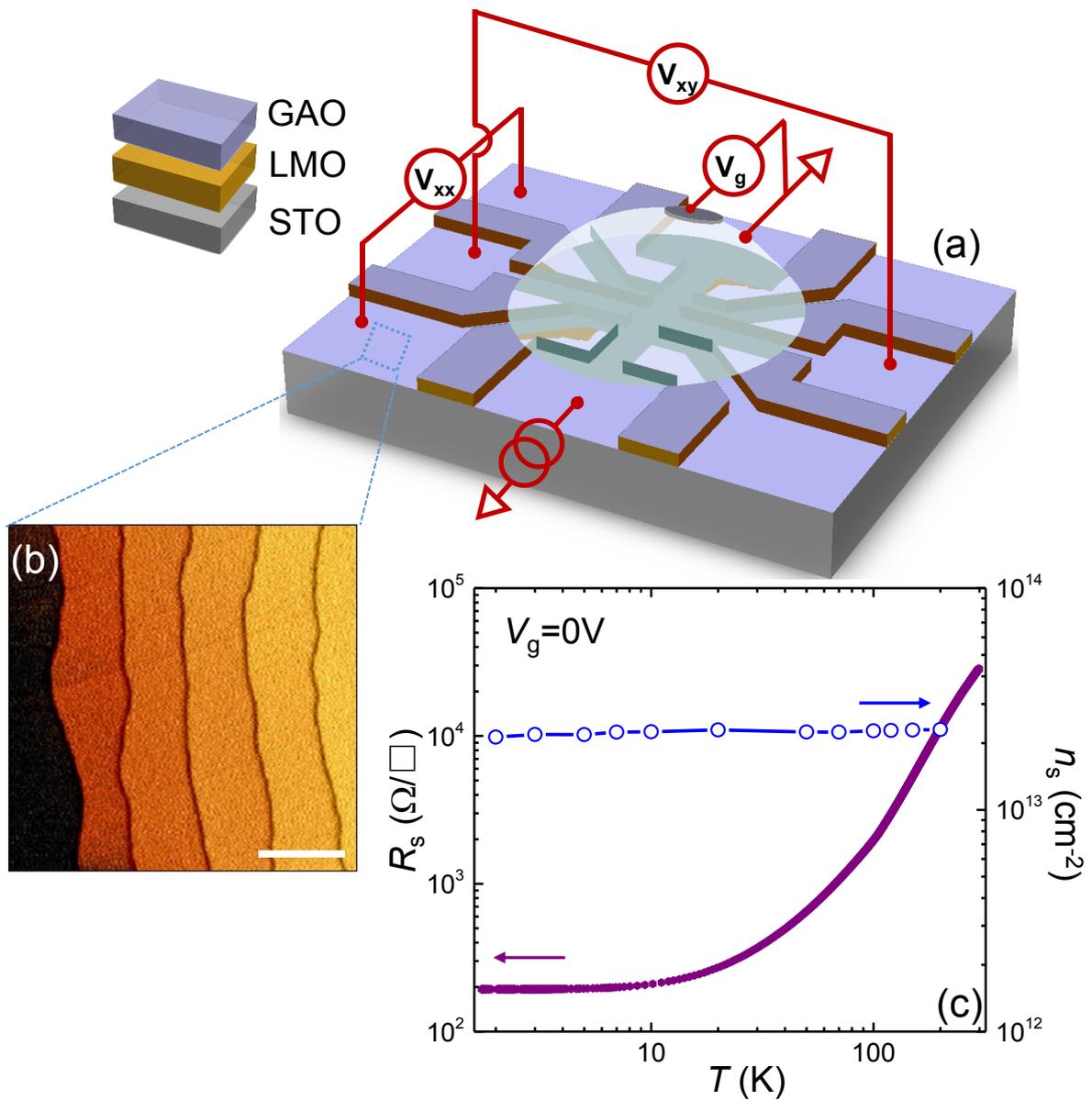

**Figure 1.** (a) A sketch of the EDLT configuration for the ionic-liquid-gated GAO/STO Hall bar interface; (b) A typical AFM image measured in the region of the Hall bar channel for a 2.5 uc GAO deposited on patterned STO substrate. The scale bar is 500 nm. (c) Transport properties of GAO/STO interface with a GAO thickness of 2.5 uc before dropping the ionic liquid.

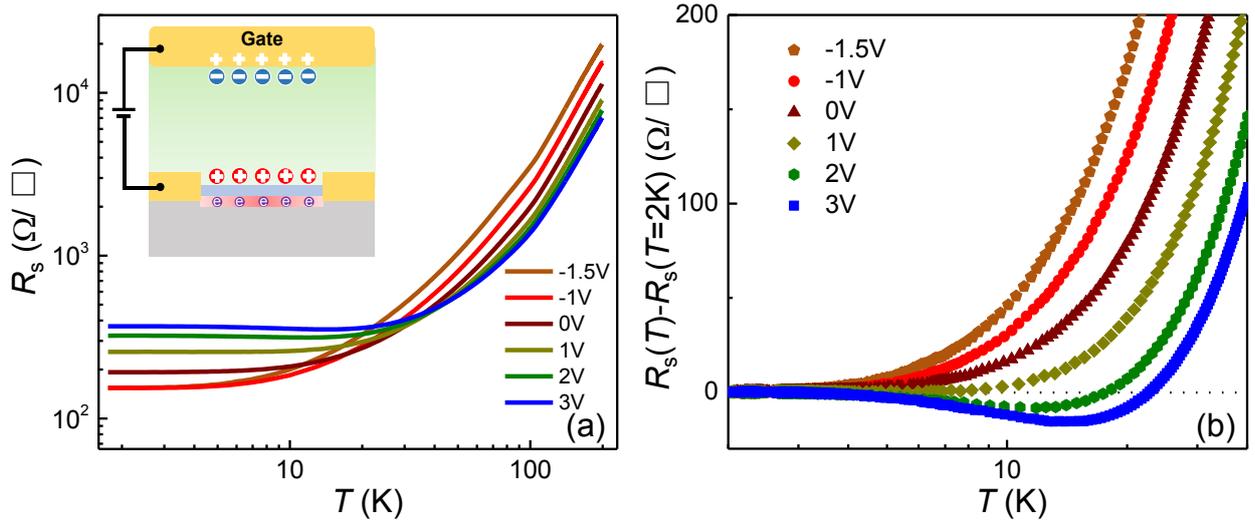

**Figure 2**. Field effect modulation of transport properties of the GAO/STO interface by EDLT. (a) Temperature dependent sheet resistance at various gate voltages. The inset shows the basic mechanism of EDLT under applied positive voltages (b) Normalized sheet resistance $R_s(T)$-$R_s$(2 K) at low temperature between 2 K and 40 K, which shows strong signature of Kondo effect.

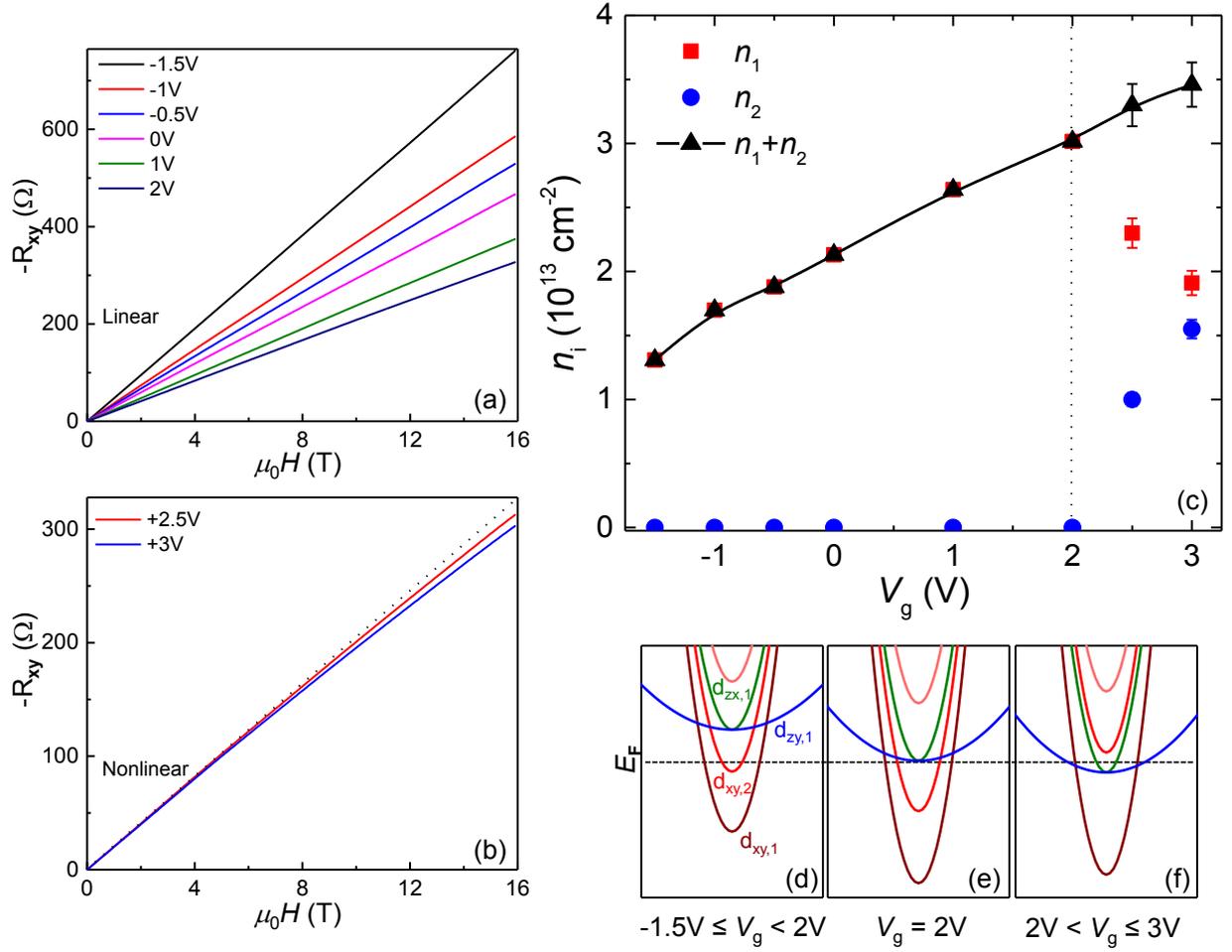

**Figure 3**. Observation of Lifshitz transition in transport at GAO/STO interface and its energy bands origin. (a) Measured Hall resistance versus magnetic field for -1.5 V ≤ $V_g$ ≤ 2 V, at T=2 K. (b) Magnetic field dependence of Hall resistance for 2 V < $V_g$ ≤ 3 V, at T=2 K. The black dashed line is linear which is guides to the eye. A transition is observed between two different types of magnetic field dependencies at $V_g$=2 V, from linear to nonlinear behavior, which is the characteristic feature of Lifshitz transition. (c) Carrier density versus gate voltage measured at 2 K. The dashed line indicates the Lifshitz carrier density, $n_L$=3×10$^{13}$ cm$^{-2}$. (d)-(f) Schematic band structures of GAO/STO interface under -1.5 V≤ $V_g$ <2 V, $V_g$=2 V and 2 V< $V_g$ ≤2 V, respectively.

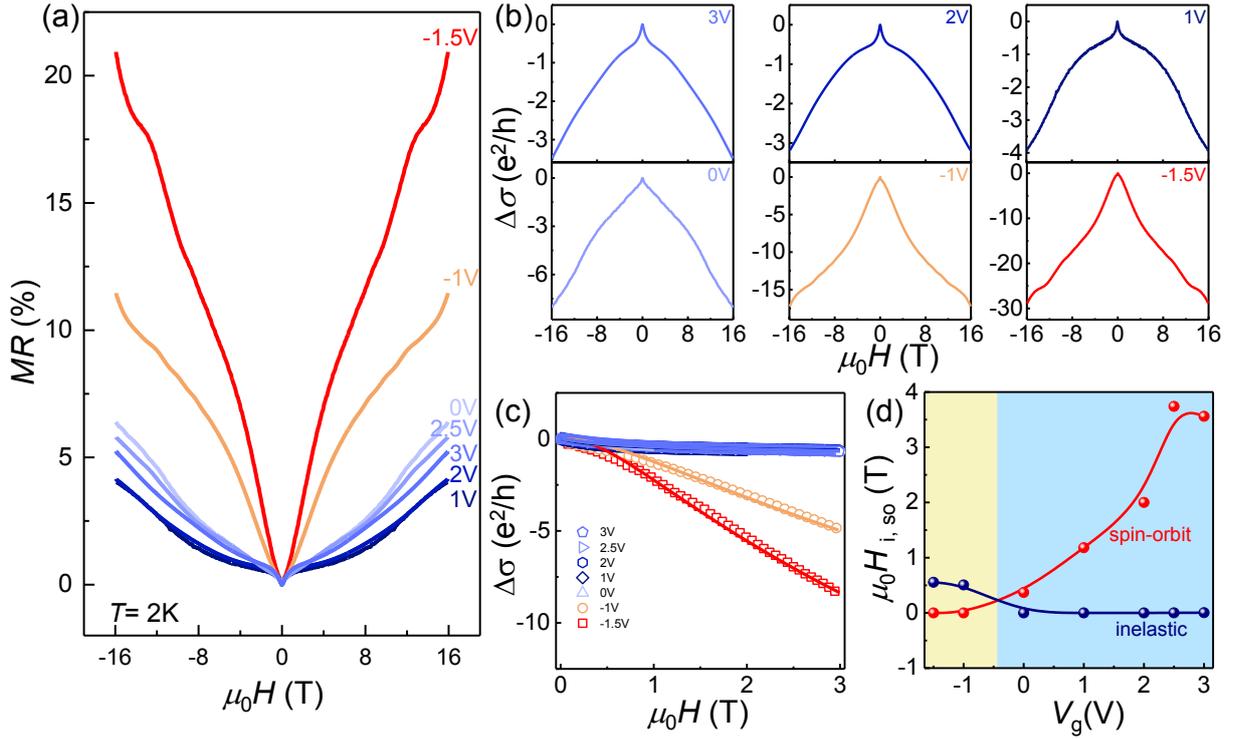

**Figure 4.** Tunable Rashba spin-orbit interaction at GAO/STO interface modulated by ionic liquid assisted gate voltages. (a) Magnetoresistance $MR=[R(\mu_0H)-R(0)]/R(0)\times100\%$ at various gate voltages. (b) Normalized conductivity $\Delta\sigma=\sigma(\mu_0H)-\sigma(0)$ in units of quantum conductance ($e^2/h$, e is the unit charge and h is the Planck constant) as a function of magnetic fields at different $V_g$. The cusps near zero magnetic field is the typical evidence of WAL. (c) Fitting the WAL effect according to the Maekawa-Fukuyama theory. The solid lines are the fitting results. (d) Gate voltage dependence of the contribution of spin-orbit interaction and inelastic scattering.

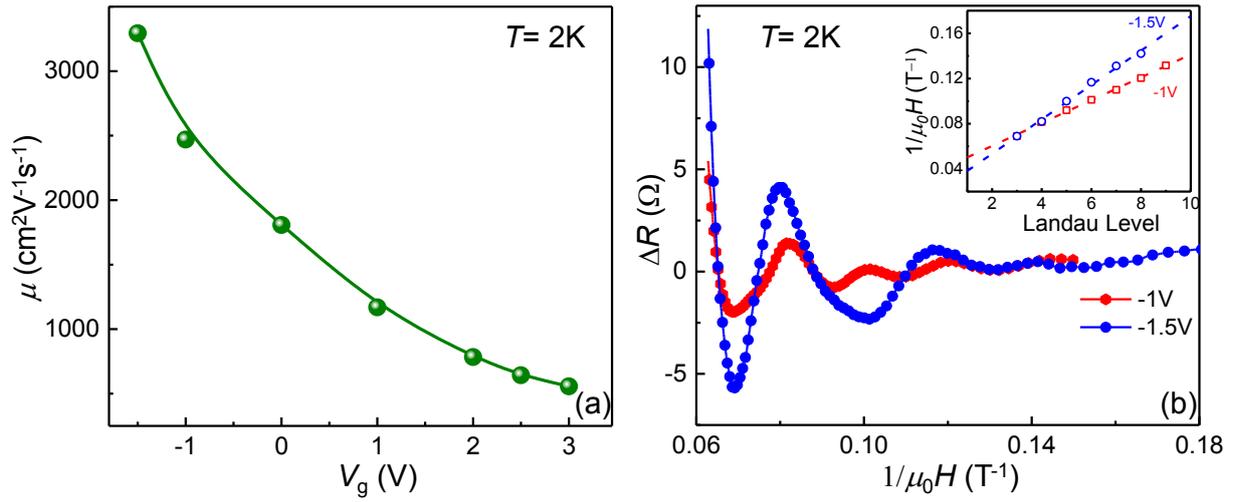

**Figure 5.** Enhanced mobility and quantum oscillations modulated by ionic liquid gating at the GAO/STO interface. (a) Mobility as a function of the applied $V_g$, measured at 2 K. (b) Amplitude of the SdH oscillation versus the reciprocal magnetic field. The inset shows the position of the oscillation peak in $1/\mu_0 H$ versus the effective Landau level.

**TOC Graphic:**

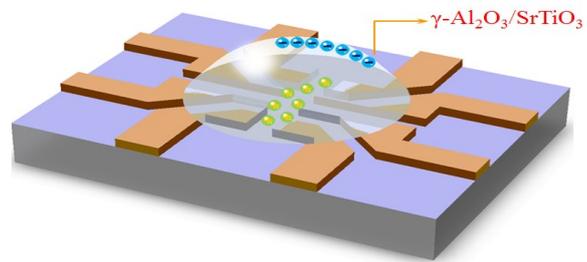